\documentclass[10pt]{iopart}  
 
\usepackage{graphicx}
\usepackage{bm}
\usepackage{color}
\usepackage[square,sort&compress,numbers]{natbib}
\usepackage[bookmarks,colorlinks,allcolors=blue]{hyperref}
 
\begin{document} 

\title[Beaming electron firehose instability]{A firehose-like aperiodic instability of the counter-beaming electron plasmas}

\author{R.~A. L\'{o}pez$^{1}$, M. Lazar$^{1,2}$, S.~M. Shaaban$^{1,3}$, S. Poedts$^{1}$ and P.~S. Moya$^{1,4}$} 
\address{$^{1}$Centre for mathematical Plasma Astrophysics, KU
  Leuven, Celestijnenlaan 200B, B-3001 Leuven, Belgium}
\address{$^{2}$Institut f\"{u}r Theoretische Physik, Lehrstuhl IV:
  Weltraum- und Astrophysik, Ruhr-Universit\"{a}t Bochum, D-44780
  Bochum, Germany}
\address{$^{3}$Theoretical Physics Research Group, Physics
  Department, Faculty of Science, Mansoura University, 35516,
  Mansoura, Egypt}
\address{$^{4}$Departamento de F\'{\i}sica, Facultad de Ciencias, Universidad de Chile, Santiago, Chile}
\ead{rlopez186@gmail.com} 
 
\begin{abstract} 
Depending on the physical conditions involved the beam plasma systems
may reveal new unstable regimes triggered by the wave instabilities of
different nature. We show through linear theory and numerical
simulations the existence of an aperiodic electromagnetic instability
which solely develops and control the stability of two symmetric plasma populations counter-moving along the regular magnetic field with a relative drift, $v_d$, small enough to not exceed the particle thermal speed, $\alpha_e$. Emerging at highly oblique angles this mode resembles properties of the aperiodic firehose instability driven by temperature anisotropy.
The high growth rates achieved with increasing the relative drift or/and decreasing the plasma beta parameter lead to significant saturation levels of the fluctuating magnetic field power, which explain the relative fast relaxation of electrons. 
For $v_d>\alpha_e$ this instability can coexist with the electrostatic two-stream instability, dominating the long-term dynamics of the plasma as soon as $v_d$ has relaxed to values smaller than the thermal speed. 

\end{abstract}

%
\vspace{2pc}
\submitto{\PPCF}
%
\maketitle
%
\ioptwocol

\section{Introduction} \label{sec:intro}

Either magnetized or not plasma systems are highly susceptible to aperiodic instabilities of electromagnetic fields \cite{Camporeale2008, Remya2013, Hellinger2018, Lopez2019, Palodhi2019}. These are zero-frequency ($\omega = 0$) waves with spatial propagation, i.e., with finite wave-numbers ($k \ne 0$), but locally their amplitude is purely growing in time with a rate ($\gamma > 0$), usually much higher than that of the periodic modes. In unmagnetized plasmas notorious are the so-called Weibel or magnetic instabilities, which can be induced by the kinetic anisotropies of plasma populations, e.g., temperature anisotropy or counter-beaming populations \cite{Weibel1959, Fried1959}, and are frequently invoked to explain the origin of cosmological magnetic field seeds, e.g., in the early Universe \cite{Schlickeiser2005, Lazar2009}, and the filamentation of energetic plasma beams \cite{Stockem2008,Bret2009,Palodhi2019}. However, the influence of a guiding stationary magnetic field on filamentation instability is not clear yet, Vlasov and particle-in-cell (PIC) simulations showing contradictory results \cite{Stockem2008,Palodhi2019}. Instead, the aperiodic mirror~\cite{Remya2013, Noreen2017, Shaaban2018} and firehose instabilities~\cite{Li2000, Camporeale2008, Shaaban2019} may  develop efficiently in finite beta plasmas, constraining any anisotropic temperature \citep{Stverak2008, Bale2009, Shaaban2018} induced by magnetic compression or adiabatic expansion along the magnetic field lines (e.g., solar outflows in the heliosphere). 

Counter-beaming plasma systems are of particular interest in astrophysical and experimental setups \cite{Fox2013,Che2014,Zhang2018,Du2019}, i.e., in fusion and plasma experiments the interest is to avoid the formation of escaping beams and stabilize plasma systems, while in astrophysics plasma beams are widely invoked, likely, at the origin of various emissions and fluctuations which trigger their relaxation \cite{Vocks2005, Umeda2006,Che2019}. Energetic beams with speed exceeding the mean thermal speed are highly susceptible to electrostatic instabilities~\cite{Lapenta2007}, and the electromagnetic modes may only hardly compete in nonrelativistic conditions~\citep{Lazar2009}. Less energetic beams with drifting (or beaming) speed lower than thermal speed have not been explored in detail in the past. Such beams guided by the magnetic fields lines can be associated with the incipient beaming formation in plasmas \citep{Marsch2006,Tautz2011, Skoutnev2019}. In this case the electrostatic instability does not easily develop, but the full wave-vector spectrum of electromagnetic (EM) fluctuations may unveil electromagnetic growing modes expected to destabilize the more or less symmetric plasma beams~\cite{Che2019}. Better known are probably the regimes of asymmetric beams, such as the electron strahls observed in the solar wind, less dense but hotter than core electrons. These strahls are responsible for the main electron heat flux, and can be destabilized either by the whistler heat-flux instability, highly conditioned by their thermal asymmetry, or by the firehose heat flux instability, if beaming speed exceeds the thermal speed of the strahl \citep{Shaaban2018b,Lopez2019b,Tong2019a,Tong2019b}. In the later case, the interplay with the electrostatic instabilities predicted by linear theory must also be investigated.

In the present paper, using linear theory and Particle-in-Cell (PIC) simulations, we show that in a magnetized plasma the field-aligned counter-beaming populations of electrons can excite an aperiodic instability with linear properties similar to the aperiodic electron firehose instability (AEFHI) driven by anisotropic temperatures $T_\parallel > T_\perp$, where $\parallel$ and $\perp$ denote directions with respect to the magnetic field. A parallel can be made with the Weibel and filamentation instabilities \citep{Weibel1959, Fried1959} in unmagnetized plasmas, one driven by temperature anisotropies (Weibel) and the other one induced by counter-beaming plasmas, but both triggered by the same mechanism and both showing similar properties. In order to demonstrate the same similarity between firehose-like instabilities, here we assume symmetric counter-beams, i.e., with the same densities, temperatures and counter-beaming speeds, and isolate from electrostatic competitors by considering beaming speeds less than thermal speeds. However, for a complete parametrization certain limit conditions are also described, allowing us to understand how this instability may convert, e.g., for less symmetric or more energetic beams, and connect to other instability conditions already described in the literature.
The results are discussed in the next section, starting with an extended linear analysis of the instability growth rates and their variations with the angle of propagation and the main plasma parameters. The analysis is completed with insights from PIC simulations, which allow us to understand the long term evolution of growing fluctuations and their back reaction on particles, contributing to the relaxation of counter-beams. Possible interplay and competition of the aperiodic beaming firehose instability with the electrostatic two-stream instability (ETSI) is also discussed. In the last section we summarize our main results and present conclusions.

\section{Results} \label{sec:results}

\subsection{Linear theory}
%
%
Our model consist in electron-proton plasma composed by protons and two counter-beaming electron populations (in a frame fixed to protons). The electron
velocity distribution function (VDF) is written as
\begin{equation}
  f_e\left({ v_{\perp},v_{\parallel}}\right) =
  \frac{n_{1}}{n_0}~f_{1}\left({ v_{\perp},v_{\parallel}}\right)
  +\frac{n_{2}}{n_0}~f_{2}\left({ v_{\perp},v_{\parallel}}\right)\,,
\end{equation}
where $n_0$ is the total electron number density (equal to the proton
density $n_0=n_p$), $n_j$ is the number density, and $f_j$ the velocity
distribution function, of the $j$-th beam. Each individual beam
distribution function is an isotropic drifting Maxwellian of the form
\begin{equation}
f_j(v_\perp,v_\parallel)=\frac{1}{\pi^{3/2}\alpha_j^3}
\exp\left\{-\frac{v_\perp^2}{\alpha_j^2}
-\frac{(v_\parallel-U_j)^2}{\alpha_j^2}\right\}\,,
\end{equation}
where $\alpha_j=(2k_BT_j/m_e)^{1/2}$ is the thermal velocity and $U_j$
the drift velocity of the $j$-th beam. Using the zero current
condition, the drift velocities are related by
$U_2=-n_1/n_2\,U_1$. On the other hand, protons are described by an isotropic non-drifting Maxwellian VDF, with the same temperature as electrons.

We start the present analysis by considering two symmetric counter-beaming electron populations. Then $n_1=n_2=0.5n_0$, $|U_1|=|U_2|=v_d$, and
$\alpha_1=\alpha_2=\alpha_e$. The plasma beta is calculated
considering the total electron density $\beta_j=8\pi
n_0k_BT_j/B_0^2$, and the plasma to gyro-frequency ratio is
$\omega_{pe}/\Omega_e=20$. For a fixed value of the electron plasma beta, $\beta_e$, we analyze the full spectrum of unstable modes triggered by the relative drift of electron populations. To do so, using the dispersion solver developed by~\cite{Lopez2017,Lopez2019}, we solve the general linear dispersion relation for arbitrary angle of propagation.

\begin{figure*}
  \begin{center} 
    \includegraphics[width=0.9\textwidth]{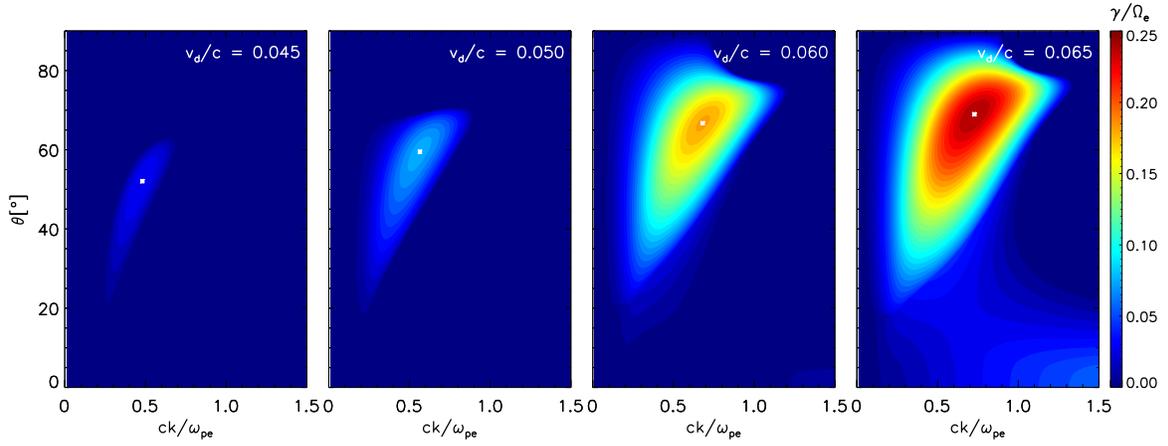}
    \caption{\label{beta2}Linear growth rates for BEFHI,
      $\gamma/\Omega_e$, for $\beta_e=2$ and various drift velocities, $v_d/c=0.045$, $0.05$, $0.06$, and $0.065$.}
  \end{center}
\end{figure*}

In figure~\ref{beta2} we present the growth rate, $\gamma/\Omega_e$,
obtained for $\beta_e=2.0$ and various drift velocities, increasing from left to right, $v_d/c=0.045$, $0.05$, $0.06$, and $0.065$, as a function of angle of propagation, $\theta$, and normalized wave number, $ck/\omega_{pe}$, where $c$ is the speed of light. Here we have used the same color bar for all the panels. In each panel, the fastest growing mode is marked with a white asterisk. Starting from the left, for $v_d/c=0.045$, the fastest growing mode is located at $ck/\omega_{pe}\approx0.48$ and $\theta\approx52^\circ$, with a relatively small maximum growth rate $\gamma_{\rm max}/\Omega_{e}\approx0.034$. This is an oblique and purely aperiodic unstable mode, with zero real frequency, $\omega_r=0$ (not shown here).  The rest of the panels clearly show that this unstable mode is markedly stimulated by the increase in the drift velocity. In the second panel, for $v_d/c=0.05$, the fastest growing mode develops at higher angle and wave number, $\theta\approx59.4^\circ$ and $ck/\omega_{pe}\approx0.56$, with a growth rate of $\gamma_{\rm max}/\Omega_{e}\approx0.08$. Third panel, for $v_d=0.06$, shows higher growth rates, with the fastest growing mode of $\gamma_{\rm max}/\Omega_{e}\approx0.19$, located at $\theta\approx66.6^\circ$ and $ck/\omega_{pe}\approx0.68$. Finally, the last panel for $v_d/c=0.065$ shows that higher values of drift velocity trigger the quasi-parallel ETSI, that can be seen for high wave numbers (around $ck/\omega_{pe}\approx 1.0$ and small angles of propagation, but with very low growth rates for the range under consideration, $\gamma/\Omega_{e}<0.1$. For this case the fastest growing mode still comes from the oblique instability, located at $\theta\approx68.9^\circ$ and $ck/\omega_{pe}\approx0.72$ and with $\gamma_{\rm max}/\Omega_{e}\approx0.25$. This unstable aperiodic mode is qualitatively very similar to the AEFHI driven by electron temperature anisotropy $T_\parallel>T_\perp$, see figure~5 in~\cite{Shaaban2019}. We see that they agree in the range of unstable wave numbers and propagation angles, and as we will discuss later, they are trigger by similar mechanisms. Is for this reason that from now on we will name this instability as Beaming electron Firehose-like instability (BEFHI).

\begin{figure*}
  \begin{center} 
    \includegraphics[width=0.9\textwidth]{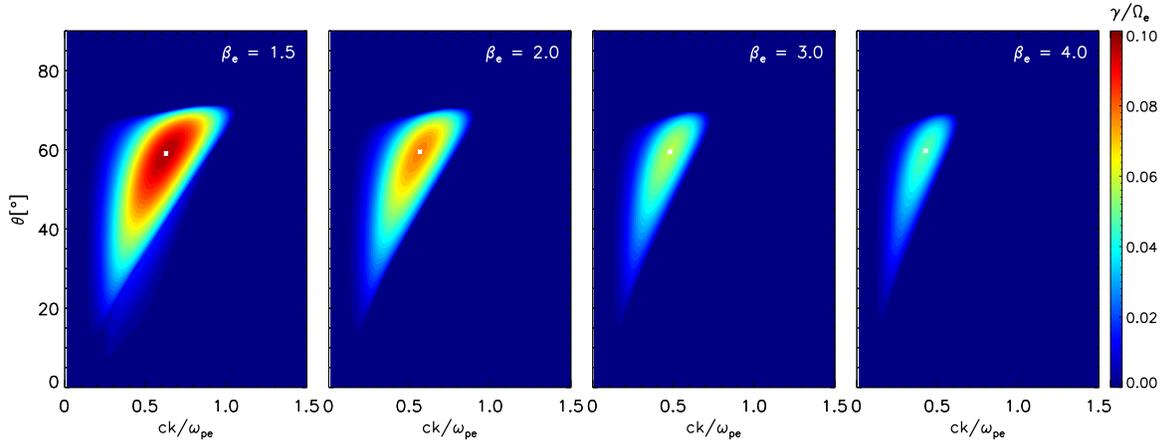}
    \caption{\label{vd005}Linear growth rates for BEFHI,
      $\gamma/\Omega_e$, for $v_d/c=0.05$ and various electron plasma beta, $\beta_e=1.5$, $2.0$, $3.0$, and $4.0$.}
  \end{center}
\end{figure*}

In figure~\ref{vd005} we study the influence of electron temperature, or the plasma beta parameter  $\beta_e=1.5$, $2.0$, $3.0$ and $4.0$ on the BEFHI, for a fixed drift velocity, $v_d/c=0.05$. In this case is clear that the instability is inhibited by increasing plasma beta. The maximum growth rate for all cases is obtained for the lowest beta, $\beta_e=1.5$. The angle of propagation of the fastest growing mode is not affected by the increase of plasma beta, being around $\theta\approx59^\circ$ for all cases. On the other side, the wave number of the fastest growing mode decreases as plasma beta increases, $ck/\omega_{pe}\approx0.62$, $0.56$, $0.47$, and $0.42$, from left to right, respectively. At the same time the maximum growth rate decreases as $\gamma_{\rm max}/\Omega_{e}\approx0.1$, $0.08$, $0.059$, and $0.049$.


\begin{figure} 
  \begin{center} 
    \includegraphics[width=0.5\textwidth]{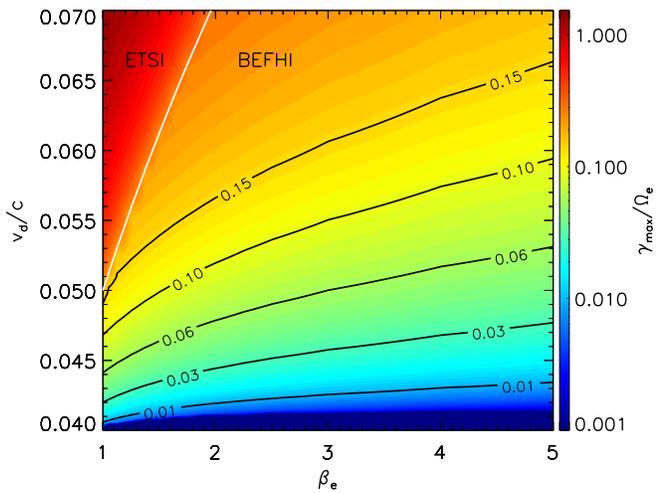}
    \caption{\label{gmax} Maximum growth rate as a function of plasma beta and drift velocity, $\beta$ and $v_d/c$. The color palette is in logarithm scale. In white we show the contour for $v_d=\alpha_e$.}
  \end{center}
\end{figure}

The results of our linear analysis are summarized in figure~\ref{gmax}. Here we show the maximum growth rates, including those from figures~\ref{beta2} and \ref{vd005}, as a function of electron plasma beta and drift velocity, $\beta_e$ vs. $v_d/c$, with the color palette in logarithm scale. We have also highlighted some thresholds values in black, $\gamma_{\rm max}/\Omega_e=0.01$, $0.03$, $0.06$, $0.1$ and $0.15$. When extracting the maximum growth rate, we can not avoid the contributions of the ETSI, which as expected, is dominant for drift velocities satisfying $v_d>\alpha_e$. The white line in figure~\ref{gmax} shows the contour $v_d/c = \sqrt{\beta_e}\,(\Omega_e/\omega_{pe})$, i.e., $v_d=\alpha_e$. As expected, for drift velocities higher than thermal velocity (above the white line), the dominant growth rates are given by the ETSI (dark red contours). However, it is important to mention that this does not mean that the BEFHI become stable above the $v_d=\alpha_e$ line. Thus, depending on the plasma parameters the ETSI and the BEFHI may coexist and compete. 
%
\subsection{PIC Simulations}

In order to study the evolution of the BEFHI and its
competition with quasi-parallel modes, we ran a series of 2.5D PIC
simulation, adapted from the KEMPO1 explicit code of Matsumoto and
Omura~\cite{Matsumoto1993}. We use a spatial grid of $n_x\times
n_y=1024\times1024$, with $400$ particles per species per grid
cell. The box length is $L_x=L_y=307.2\,c/\omega_{pe}$, with the cell
width $\Delta x=\Delta y=0.3\,c/\omega_{pe}$. We use the real mass
ratio $m_p/m_e=1836$, and the plasma to gyro-frequency ratio
$\omega_{pe}/\Omega_e=20$. The background magnetic field is in the $x$
direction, $\mathbf{B}_0=B_0\,\mathbf{\hat{x}}$. Finally, the time step is $\Delta t=0.01/\omega_{pe}$, and the simulation runs until
$t_{\rm max}=81.92/\Omega_e$. 
%
\begin{table}
    \caption{Initial plasma parameters for the simulation.\label{t1}}
    \footnotesize
    \begin{tabular}{@{}llll}
    \br
    Parameter & Case I  & Case II & Case III\\
    \mr
        $\beta_e$ & $4.0$ & $2.0$ & $1.0$ \\
        $v_d/c$ &  $0.07$ & $0.06$ & $0.06$\\
        $v_d/\alpha_e$ &  $0.70$ & $0.85$ & $1.20$\\
        $\gamma_{\rm max}/\Omega_e$ & $0.21$ & $0.19$ & $1.91$ \\
        \br
    \end{tabular}\\
\end{table}
\normalsize
%
To save computational resources, we have run cases with large growth rates, so the instabilities can develop in a shorter temporal window. The three cases selected are listed in Table~\ref{t1}. 

\begin{figure}[h!]  
  \begin{center} 
    \includegraphics[width=0.45\textwidth]{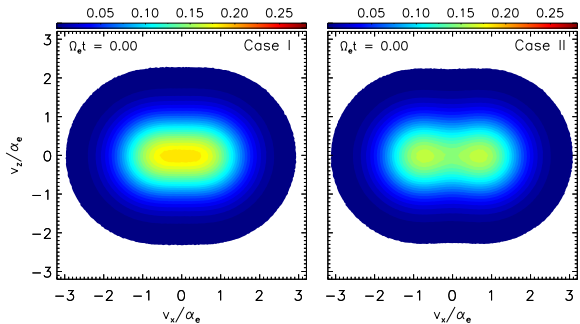}
    \caption{\label{dist}Initial electron velocity
      distribution function, $f(v_x,v_y)$, in the $v_x$--$v_z$ space, for both cases.}
  \end{center}
\end{figure}
\begin{figure}[ht!]
  \begin{center}
    \includegraphics[width=0.45\textwidth]{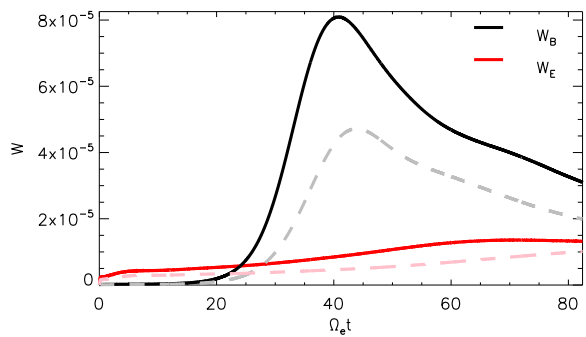}
    \caption{\label{energy}Temporal evolution of the magnetic and
      electric energy density for Case I ($v_d/c=0.07$ and $\beta_e=4.0$), and Case II ($v_d/c=0.06$ and $\beta_e=2.0$). Magnetic energy density is shown by black and grey lines, and electric energy density by red and pink, for cases I and II, respectively.}
  \end{center}
\end{figure}

We start studying two cases in which $v_d<\alpha_e$ and the BEFHI is dominant: Case I for $v_d/c=0.07$ ($v_d/\alpha_e=0.7$) and $\beta=4$, with fastest growing mode of $\gamma_{\rm max}/\Omega_e\approx0.21$; and Case II for $v_d/c=0.06$ ($v_d/\alpha_e=0.85$) and $\beta=2$, with fastest growing mode of $\gamma_{\rm max}/\Omega_e\approx0.19$. The initial total electron velocity distribution function for both cases is shown in figure~\ref{dist}, obtained from the simulation. Figure~\ref{energy} shows the temporal evolution of the magnetic energy density $W_B=\int \delta B^2/B_0^2\,dxdy$ (solid black for Case I and dashed grey for Case II) and electric energy density $W_E=\int \delta E^2/B_0^2\,dxdy$ (solid red for Case I and dashed pink for Case II), for the entire simulation period. We observe that for both cases the magnetic power, $W_B$, is dominant, increasing exponentially reaching maximum intensity around $\Omega_e t\approx40.8$ for Case I (black) and $\Omega_e t\approx43.8$ for Case II (grey). As expected, the onset time of the instability is lower for Case I, and the maximum intensity reached is higher, since it has the fastest growing mode between the two cases. On the other hand, the temporal variation of the electric energy density is very small for both cases. This is a strong evidence that the BEFHI is mainly of magnetic nature. After the saturation of the instability, the magnetic energy density undergoes a gradual decrease due to the reabsorption of the wave energy, a feature commonly observe in the evolution of the electron firehose instability~\cite{Yoon2017,Lopez2019}.
\begin{figure}[h!]
  \begin{center}
    \includegraphics[width=0.45\textwidth]{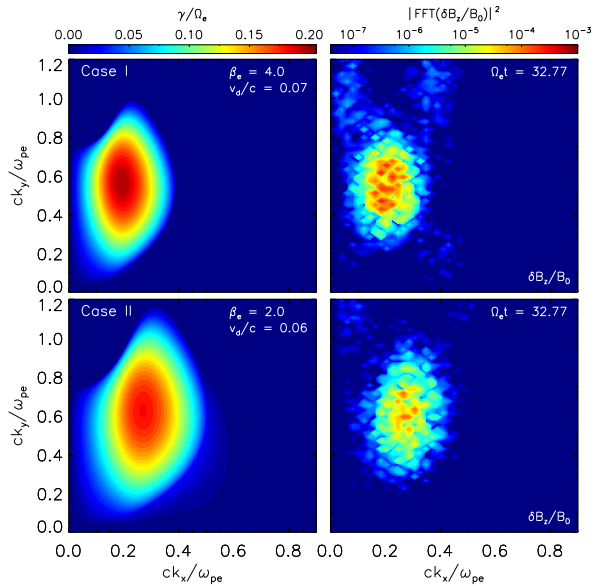}
    \caption{\label{spectra}Left panels: growth rates from linear theory. Right panel: Power spectra of the transverse magnetic field fluctuations, $\left|{\rm FFT}(\delta B_z/B_0)\right|^2$, obtained from the simulation, in logarithmic scale. Upper and lower panels correspond to Case I and Case II, respectively.}
  \end{center}
\end{figure}
In figure~\ref{spectra} we compare the prediction from linear theory for the growth rate of the BEFHI against PIC simulations. The linear growth rates are shown in left panels, for Case I on top, and Case II on the bottom panel. This time the growth rates are plotted in the $k_x$--$k_y$ plane for a better comparison between theory and simulations. On the right panels of figure~\ref{spectra} we plot the spatial power spectra of the transverse magnetic field fluctuations, $\left|{\rm FFT}\left(\delta
B_z/B_0\right)\right|^2$, obtained from the simulation, in logarithmic scale. This figure shows how the prediction from linear theory is in very good agreement with the results of our PIC simulations, confirming the existence of the electromagnetic oblique unstable mode, i.e., BEFHI. 
The aperiodic nature of the BEFHI is shown in figure~\ref{fields}, where we plot three snapshots of the transverse magnetic field fluctuations in the space domain ($\delta B_z(x,y)/B_0$), for Case I (top panels in figure~\ref{spectra}), at times at $\Omega_e t=20.48$ (left), $\Omega_et=30.21$ (middle), and $\Omega_et=40.45$ (right). We observe that these fluctuations are mainly oblique to the background magnetic field and do not propagate in space, but only grow in time.\\
\begin{figure}[h!]
  \begin{center}
    \includegraphics[width=0.5\textwidth]{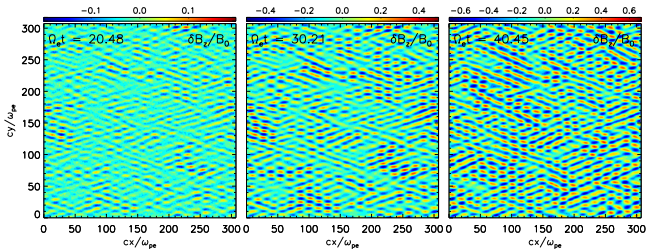}
    \caption{\label{fields}Snapshots of the transverse magnetic field
      fluctuations, $\delta B_z/B_0$, as a functions of space, for Case I.}
  \end{center}
\end{figure}
Finally, figure~\ref{fx} shows the reduced eVDF in parallel direction, $f(v_x)$, for different stages of Case I (left) and Case II (right) simulations. For Case I, initially at $\Omega_et=0$, it is not possible to distinguish the two peaks from each beam. However, due to the drift between both beams we observe a flat-top distribution that then evolves to a more isotropic distribution, with a reduced drift, so the core of the resultant distribution increases its density resembling a single Maxwellian eVDF. The reduction in the drift velocity is more clearly shown in the left panel of figure~\ref{fx}, corresponding to Case II. Here it is clear that, as the BEFHI evolves in time, the initial gap between the two beams gets filled, and the core of the combined distribution increases in a similar way as in Case I.  
\begin{figure}[h!] 
  \begin{center}
    \includegraphics[width=0.45\textwidth]{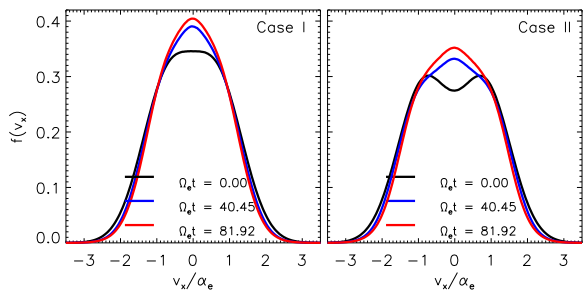}
    \caption{\label{fx}Reduced eVDF in parallel direction, $f(v_x)$, for different stages of the two cases, Case I (left) and Case II (right).}
  \end{center}
\end{figure}
%

\subsection{Competition with electrostatic two-stream instability}

\begin{figure} 
  \begin{center} 
    \includegraphics[width=0.45\textwidth]{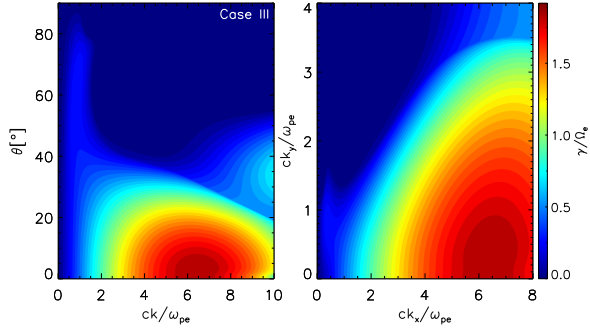}
    \caption{\label{disp_case3}Linear growth rates for BEFHI,
      $\gamma/\Omega_e$, for Case III ($\beta_e=1$ and $v_d/c=0.06$). Left panel: $ck/\omega_{pe}$ vs $\theta$. Right panel: $ck_x/\omega_{pe}$ vs. $ck_y/\omega_{pe}$.}
  \end{center}
\end{figure}

As it is shown in figures~\ref{beta2} and \ref{gmax}, as the drift velocity of the beams increases the quasi-parallel aperiodic two-stream instability (ETSI) become stronger,
being the dominant mode for $v_d/\alpha_e>1$. To explore this regime, in Case III we select a configuration in which $v_d/\alpha_e = 1.2$, with the BEFHI still present but about one order of magnitude weaker. Figure~\ref{disp_case3} shows the growth rate for this case. In the left panel we plot $ck/\omega_{pe}$ vs. $\theta$, showing that the fastest growing mode is at high wave numbers and in parallel direction, with a growth rate of $\gamma_{\rm max}/\Omega_e\approx1.91$ at $\theta=0^\circ$ and $ck/\omega_{pe}\approx6.5$. We can observe the weak signature of the BEFHI at very high angles and low wavenumbers, where the growth rate is around $\gamma_{\rm max}/\Omega_e\approx0.31$, which is still higher than the previous two cases. Right panel of figure~\ref{disp_case3} shows the same information but in the $k_x$ vs. $k_y$ plane (to compare with simulations). We ran a simulation for this case, using the same configuration described before, except that we have refined the grid to $\Delta x=\Delta y=0.1\,c/\omega_{pe}$, so we can resolve small scale fluctuations, appearing at large wavenumbers. Figure~\ref{spectra_case3} shows the power spectra of both parallel electric and transverse magnetic field fluctuations, $\left|{\rm FFT}(\delta E_x/B_0)\right|^2$ (top) and $\left|{\rm FFT}(\delta B_z/B_0)\right|^2$ (bottom), respectively, for two different times of the Case III simulation. Left panels show the situation at $\Omega_et=2.05$ in which it can be seen that the ETSI and BEFHI modes coexist. The power in the parallel electric field (top left) shows the presence of mainly ETSI quasi-parallel modes around $ck/\omega_{pe}\approx6$, in agreement with linear theory shown in figure~\ref{disp_case3} right panel. At the same time, transverse magnetic field fluctuations (bottom left) show a very weak presence of the BEFHI modes at highly oblique angles around $ck_y/\omega_{pe}\approx1$, again consistent with the picture of figure~\ref{disp_case3}. 

\begin{figure} 
  \begin{center}
    \includegraphics[width=0.45\textwidth]{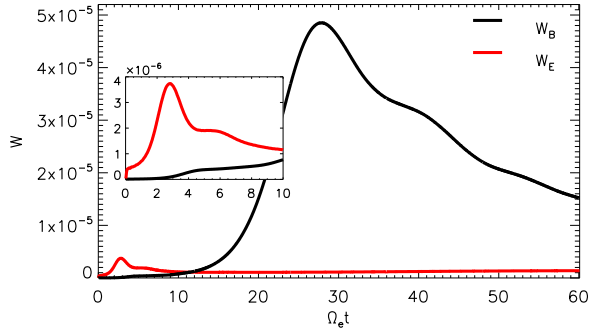}
    \caption{\label{energy_case3}Temporal evolution of the magnetic and
      electric energy density for Case III.}
  \end{center}
\end{figure}

\begin{figure}[h!]
  \begin{center} 
    \includegraphics[width=0.45\textwidth]{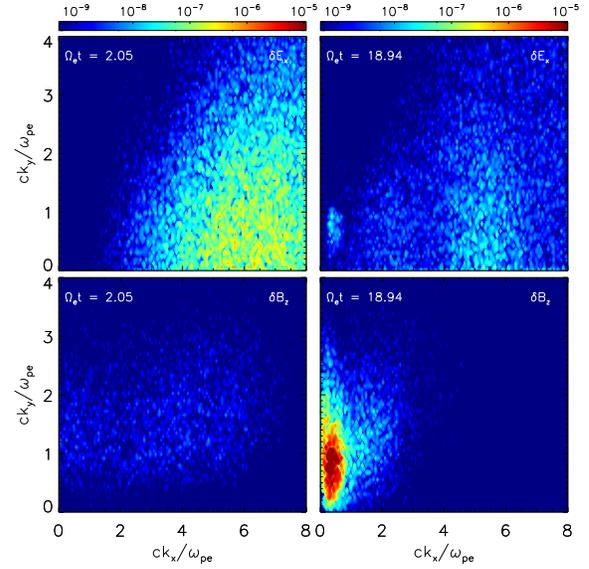}
    \caption{\label{spectra_case3}Power spectra of the parallel
      electric and transverse magnetic field fluctuations,
      $\left|{\rm FFT}(\delta E_x/B_0)\right|^2$ (top) and
      $\left|{\rm FFT}(\delta B_z/B_0)\right|^2$ (bottom), respectively,
      obtained for different stages of the simulation for Case III.}
  \end{center}
\end{figure}

\begin{figure}[h!] 
  \begin{center}
    \includegraphics[width=0.4\textwidth]{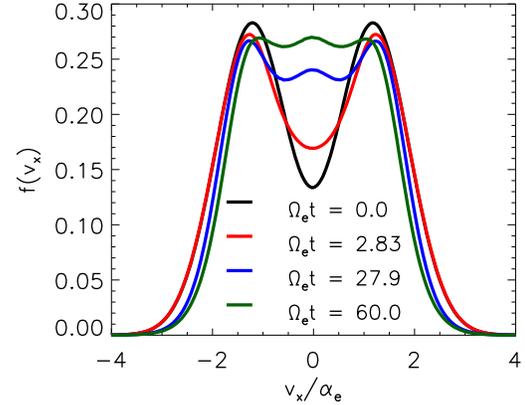}
    \caption{\label{fx_case3}Reduced eVDF in parallel direction, $f(v_x)$, for different stages of Case III.}
  \end{center}
\end{figure}

\begin{figure} 
  \begin{center}
    \includegraphics[width=0.5\textwidth]{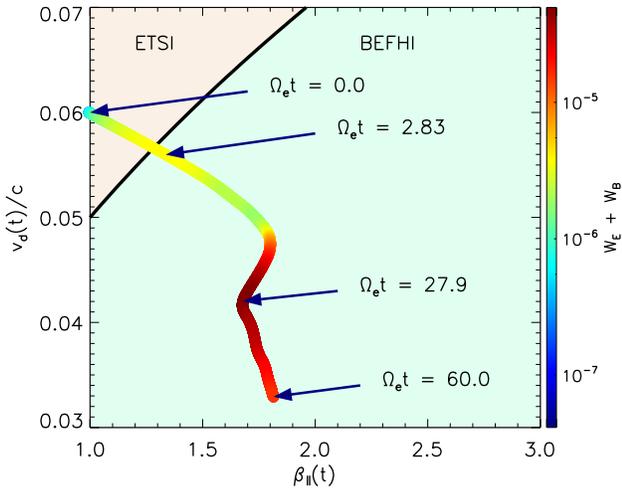}
    \caption{\label{path_case3}Dynamical path of the simulation for Case III, $v_d(t)/c$ vs. $\beta_\parallel(t)$. Color palette indicates the total electromagnetic energy density in logarithm scale. Black line shows the condition $v_d=\alpha_e$, as in Fig.~\ref{gmax}.}
  \end{center}
\end{figure}

The time evolution of both instabilities can be seen in figure~\ref{energy_case3}, where we show the electric (red) and magnetic (black) field energy density as function of time during Case III simulation. Although the ETSI is dominant according to the linear theory, it can be seen that the electric energy grows faster but saturates at lower levels than the magnetic energy density. At $\Omega_et=2.05$, during the linear stage of the ETSI, the electric energy density is dominant in both Figs.~\ref{spectra_case3} and \ref{energy_case3}. However, this instability is rapidly inhibited, saturating around $\Omega_et\approx2.83$. At the same time but with a slower rate, the magnetic energy density exponentially grows, saturating around $\Omega_et\approx27.9$ with a level of almost one order of magnitude higher than the electric energy density. This magnetic instability corresponds to the BEFHI. The spectral power showed in right panels of figure~\ref{spectra_case3}, at $\Omega_et=18.94$, is concentrated in the transverse magnetic field at highly oblique angles (around $ck_x/\omega_{pe}\approx0.3$ and $ck_y/\omega_{pe}\approx0.8$), which is again consistent with linear theory predictions (see figure~\ref{disp_case3} right panel).
The fast inhibition of the ETSI and the later dominance of the BEFHI can be explained by looking at the evolution of the eVDF and its moments, as it is shown in figures~\ref{fx_case3} and \ref{path_case3}. Figure~\ref{fx_case3} shows the evolution of the reduced eVDF along the magnetic field direction at different moments of Case III simulation. The initial configuration (in black) correspond to two clearly separated beams unstable to the ETSI. As the simulation evolves the gap between each peak is filled, which means that the drift velocity is decreasing. By the time the ETSI has saturated, $\Omega_et=2.83$, is still possible to distinguish each beam, and the drift velocity is still significant (red line), so the distribution is still unstable. However, by this stage of the simulation the drift between the beams is smaller than the thermal speed, so the BEFHI becomes the dominant instability. As the simulation and the BEFHI evolve, the drift between both beams decreases giving free energy to the magnetic field, that saturates at $\Omega_et=27.9$. At this time the relative speed between beams is clearly reduced, and it is difficult to distinguish to peaks in the field-aligned distribution (blue curve in Figure~\ref{fx_case3}). Then, after its saturation, the instability relaxes as some of the magnetic energy is transported back to the distribution, so that at the end of the simulation ($\Omega_et=60.0$) Figure~\ref{fx_case3}) shows a wider (hotter) single distribution. 

The situation is more clear in figure~\ref{path_case3}, where we follow the temporal evolution of the moments of the total distribution in the $v_d$ vs. $\beta$ plane. Initially the configuration is unstable to both ETSI and BEFH, but is the ETSI the dominant and the one determining the initial evolution. We observe that the simulation evolve towards the regime dominated by the BEFHI, by reducing the drift and increasing the parallel plasma beta. By the time $\Omega_et=2.83$ the simulation has crossed the threshold $v_d=\alpha_e$, to a regime where the BEFHI is dominant, which explain the reduction in the electric field energy as a consequence of the inhibition of the ETSI. After the saturation of the ETSI the electromagnetic energy density grows again, this time because of the magnetic energy increase due to the evolution of the BEFHI. Until its saturation at $\Omega_et=27.9$, the instability grows at the expense of both the drift between the beams and plasma beta. Then, as the instability relaxes (see Figure~\ref{energy_case3}) and the magnetic energy density decreases, the drift gets further reduced but now plasma beta increases, showing that part of the wave energy has returned to the particles, heating the plasma.
\section{Conclusions}
\label{sec:conclusions}
 
In the present paper we have used linear theory and PIC simulations to show that in a magnetized plasma field-aligned counter-beaming populations of electrons can excite an aperiodic oblique instability (BEFHI) with linear properties similar to the aperiodic electron firehose instability (AEFHI) driven by anisotropic temperatures $T_\parallel > T_\perp$. For symmetric counter-beaming populations of electrons, i.e. with the same densities, temperatures and beaming speeds, we can associate an effective temperature anisotropy, but taking it with precaution, only for less energetic beams, where it may not prevent the competition with electrostatic instabilities. Growth rates are found to be highly sensitive to the variation of plasma parameters, increasing as the beaming speed increases (for a fixed plasma beta), and lowering with the increase of plasma beta (for a fixed beaming speed). These variations have helped us to derive and display in figure~\ref{gmax}, the numerical instability thresholds, as contours of constant, low values of maximum growth rates, in terms of drifting speed and the electron plasma beta. Theory also predicts that this instability may coexist and interplay with other beam plasma instabilities, as the case of the electrostatic two-stream instability (ETSI), that develops when beaming speed $v_d$ becomes comparable with thermal speed $\alpha$. However, BEFHI remains dominant as long as the beaming speed is less than the thermal speed. For more energetic beams with $v_d/\alpha_e>1$ the BEFHI is highly competed by the ETSI, which reaches maximum (peaking) growth rates almost one order of magnitude higher than those obtained for BEFHI. 

We confirm the linear properties and the existence of the aperiodic BEFHI beyond the linear regime through PIC simulations. First, in the $v_d/\alpha_e<1$ regime we show that the BEFHI generates purely growing magnetic field fluctuations, oscillating in space and propagating obliquely to the background magnetic field, in agreement with the linear theory predictions. The instability grows at the expense of the relative drift between the beams and subsequently the combined distribution becomes more isotropic, resembling a single Maxwellian by the end of the simulation. The simulations also show that, similar to the case of the AEFHI, the feedback between waves and particles saturates the electromagnetic energy, that then decreases while the plasma is heated~\cite{Yoon2017,Lopez2019}. 

In the simulations we have also investigated the extended regime of higher beaming speeds, i.e., for $v_d>\alpha_e$, when the ETSI and BEFHI are expected to coexist and interplay. In this regime linear theory predicts that the plasma is unstable to the ETSI, but plasma beta is large enough so the BEFHI is also present although does not dominate. For this case PIC simulations show that, indeed, the ETSI develops first reducing the drift and increasing plasma beta. However, as soon as the $v_d =\alpha_e$ threshold is reached, the electric field energy saturates and then relaxes. Meanwhile, the BEFHI grows at a slower rate, reducing drift and plasma beta, but the magnetic field energy saturates at a level about one order of magnitude higher than the electric field. Finally, after the saturation the BEFHI may also contribute to the relaxation of beams as already described for the $v_d/\alpha_e<1$ regime. 

We  expect our present results to be relevant in many specific applications of plasma beams in astrophysical and fusion scenarios, helping us to understand their time evolution and stability in magnetized plasma systems, especially at the short timescales in which beams are still accelerating or when their velocity is still smaller than the local thermal speed.

\ack 

These results were obtained in the framework of the projects SCHL 201/35-1 (DFG-German Research Foundation), GOA/2015-014 (KU Leuven), G0A2316N (FWO-Vlaanderen), and C~90347 (ESA Prodex 9). S.M. Shaaban acknowledges support by a FWO Postdoctoral Fellowship, grant No.~12Z6218N. P.S. Moya is grateful for the support of KU Leuven BOF Network Fellowship NF/19/001, and CONICyT Chile through FONDECyT Grant No. 1191351.
The computational resources and services used in this work were provided by the VSC (Flemish Supercomputer Center), funded by the Research Foundation— Flanders (FWO) and the Flemish Government—department EWI.

\bibliographystyle{iopart-num}
\bibliography{biblio}
 
\end{document}